\documentclass
{aa} % for a referee version

\catcode`\@=11 % This allows us to modify PLAIN macros.
\def\lsim{\mathrel{\mathpalette\@versim<}}
\def\gsim{\mathrel{\mathpalette\@versim>}}
\def\@versim#1#2{\vcenter{\offinterlineskip
        \ialign{$\m@th#1\hfil##\hfil$\crcr#2\crcr\sim\crcr } }}

\newcommand{\msun}{$M_\odot$} 

\usepackage{graphicx}
\begin{document}
\title{The new infra red view of evolved stars in I\,Zw\,18
%\title{A search for carbon stars in I\,Zw\,18
\thanks{Based on observations with the NASA/ESA {\it Hubble Space 
Telescope}, obtained at the Space Telescope Science Institute, 
which is operated by the association of Universities for Research 
in Astronomy, Inc., under NASA contract NAS5-26555.}}

\author{G. \"Ostlin \inst{1}
	\and	
	M. Mouhcine \inst{2,3} 
           }

\offprints{G\"oran \"Ostlin }

\institute{
	Stockholm Observatory, AlbaNova University Center, 
	SE-106 91 Stockholm, Sweden \\
	\email{ostlin@astro.su.se}
\and
	School of Physics and Astronomy, University of Nottingham, 
        Nottingham NG7 2RD
\and	
       Observatoire Astronomique de Strasbourg (UMR 7550),
       11, rue de l'Universit\'e, 67000 Strasbourg, France \\
}

   \date{Received 28/04/2004; accepted 31/10/2004}

   \abstract{
We report  results from near-infrared imaging  of I\,Zw\,18, the most 
metal-poor galaxy in the local universe  with 
NICMOS  on board the Hubble Space Telescope (HST). 
Observations were obtained in the broad F205W filter ($\sim$ K) and 
in the medium-wide bands F171M and F180M, the latter which cover strong 
molecular absorption features in cool stars.
The new data, together with previously obtained HST/NICMOS images in 
the F110W ($\sim$ J) and F160W ($\sim$ H) filters, provide a census 
of the  cool stellar population in this chemically unevolved galaxy. 
We find that stars as old as $\sim\,1\,$Gyr are required to explain the observed 
colour-magnitude diagrams. Combining broad  and medium-wide band 
photometry, we have classified the observed stars. The observed stellar 
populations, to the depth of our data, are dominated by luminous red 
supergiant stars. However, an intermediate-age component is also present. 
We have identified carbon star candidates, and  show that they 
dominate the stellar content for intermediate mass and age. We show that low metallicity 
intermediate-age stars have, for a given colour, a F110W ($\sim$ J-band) 
excess compared to local carbon stars. The carbon stars identified in 
I\,Zw\,18 are the most distant (more than 10 Mpc) resolved examples 
yet discovered.
}

\authorrunning{\"Ostlin \& Mouhcine}

\titlerunning{Carbon stars in I\,Zw\,18 }

\maketitle

\section{Introduction}
\label{intro}
There has been a strong interest in blue compact galaxies ever since
the pioneering study of I\,Zw\,18 by Searle \& Sargent (\cite{ss72}). 
The global properties of I\,Zw\,18, i.e., extremely low nebular oxygen 
abundance, $12+\log({\rm O/H}) = 7.18$ (Izotov and Thuan \cite{it99};
just a few percent of the solar abundance), very blue colours, and a 
large H{\sc i} reservoir, have been interpreted as evidence that this
galaxy may be genuinely young, presently forming its very first 
generation of stars. If its star formation history 
really is restricted to the last few 10\,Myr, the existence of 
I\,Zw\,18 demonstrates that the formation of galaxies may be continuing
also at the present cosmic  epoch. Speaking against the youth 
hypothesis is the discovery of a population of faint red stars in 
I\,Zw\,18, both in the optical (Aloisi et al. \cite{aloisi}) and in the 
near-infrared (\"Ostlin \cite{ostlin2000}). These stars have probable 
ages in the range from 100 Myr to a few Gyr, and most of these are 
expected to be stars on the asymptotic giant branch (hereafter AGB). 

The upper AGB is populated with oxygen-rich and carbon-rich stars
(i.e., AGB stars with C/O ratio smaller or larger than unity by number,
respectively).
It has been established observationally and theoretically that carbon 
star formation is more efficient  in low metallicity environments. 
The number ratio of carbon stars to M stars anti-correlates with the 
metallicity 
(Cook et al. 1986, Groenewegen \cite{groenewegen}, 
Mouhcine \& Lan\c{c}on 2003).
I\,Zw\,18 has the lowest ever measured ionised gas phase metallicity in
a galaxy (Kunth \& \"Ostlin \cite{kunth}). Thus, if there is an
intermediate-age stellar population present in I\,Zw\,18, one might expect 
 it to be dominated by carbon stars. In the same context, the 
carbon-star content of I\,Zw\,18 could provide insights into stellar evolution at low metallicity.

AGB star spectra are dominated by broad molecular  absorption bands.
Spectra of oxygen-rich stars are dominated by metal oxide bands such 
as TiO, VO, and H$_2$O. Optical and near-IR spectra of carbon stars have 
little in common
with those of oxygen-rich AGB stars apart from the CO absorption in K-band.
They are dominated by bands of CN and the Phillips \& Ballik-Ramsay 
C$_2$ systems (Barnbaum et al. 1996, Joyce 1998). Because of their high 
luminosity and distinctive spectra, carbon stars can in principle be 
identified out to great distances.
It has been argued that carbon stars have, in general, redder 
broad band colours than oxygen-rich stars. 
This statement is not entirely true. For AGB stars with low mass loss 
rate, i.e., stars with optically thin circumstellar shells, carbon stars 
display a wide range of colours, similar to the colour range spanned by 
M-stars (Alvarez et al. 2000). 
Due to CN and C$_2$ absorption in the J-band,  
the $J-K$ colour distribution of carbon stars extends to redder colours
than M-stars, i.e., $J-K\simeq\,2.5$ (Loup et al. 1998). Applying a 
sharp cutoff to separate carbon stars from oxygen-rich stars may miss 
warm carbon stars, and misclassify extreme oxygen-rich optically selected 
or dust-enshrouded stars as carbon stars. Thus, optical/near-IR colours 
are insufficient, by themselves, to discriminate carbon stars from 
oxygen-rich stars.

A more successful strategy to find carbon stars is to make use of
 spectral features that  uniquely identify such stars. Different
indices, corresponding to different spectral features, have been proposed 
in the literature (Wing 1967, Aaronson et al. 1982). The NICMOS instrument 
on the Hubble Space Telescope (HST) has a set of medium-wide filters. 
One of these, F180M in the NIC2 camera, coincides with a C$_2$ band head 
at $1.77\mu$m  (Ballik-Ramsay series, $\Delta v = 0$). 
Together with the adjacent continuum filter F171M, it can be used to 
probe the strength of the C$_2$ band head (Alvarez et al. 2000). 
%The HST/NICMOS 
%colour index $F180M-F171M$, where F171M is an adjacent 
%continuum filter, is therefore sensitive to the C$_2$ band head strength
%(Alvarez et al. 2000). 

In this paper, we report the results from HST/NICMOS medium-wide
(F171M and F180M) and broad (F205W, $\sim$ K) band imaging of I\,Zw\,18,  
which we use to investigate the nature of red stars and the upper AGB  
population. The observations and the data reductions are described in 
Sect.\,\ref{obs}. In addition we use previously obtained NICMOS data 
in the F110W ($\sim$ J) 
and F160W ($\sim$ H) filters. 
In Sect.\,\ref{surf_phot}, we discuss the surface photometry. 
The nature of the red and cool stellar content of I\,Zw\,18 identified using 
colour-magnitude and colour-colour diagrams is discussed in Sect.\,\ref{pht}. 
Conclusions are summarised in Sect.\,\ref{concl}.

\begin{figure}
 \resizebox{\hsize}{!}{\includegraphics{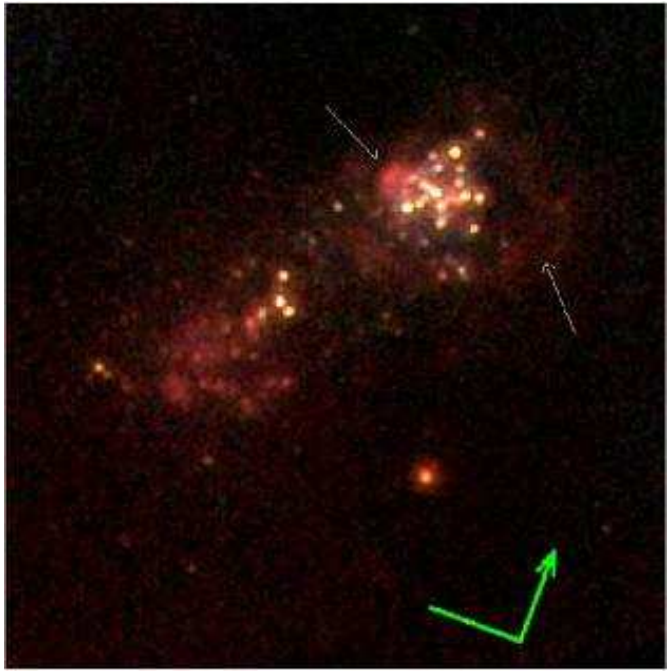}}
 \caption{RGB composite of F205W (red), F160W (green) and F110W (blue) 
 showing a field of $14\arcsec  \times 14\arcsec$. The orientation is
 indicated by the green compass in the lower right corner (north is up
 right, east is up left). The length of the bars is $2\arcsec$. Two of 
 the diffuse sources with strong excess in the F205W filter are indicated 
 by thin white arrows.}
 \label{image1}
 \end{figure}

 \begin{table}[h]
\caption[]{Log of observations for GO programme 7880, with zero points
given in the Vegamag system.} 
\begin{flushleft}
\begin{tabular}{lllr} 
\hline
Filter &  \multicolumn{2}{l}{Total exposure time} & Zero Point  \cr
\cline{2-3}
 & object  & sky  \cr
 & (s)  & (s) & \cr
\noalign{\smallskip} 
\hline 
\noalign{\smallskip}
F171M & 4629.47 & 3086.32 & 19.956 \cr
F180M & 6172.63 & 4115.09 & 19.844 \cr
F205W & 4629.47 & 3086.32 & 21.960 \cr
\hline
\end{tabular}
\end{flushleft}
\label{obstab}
\end{table}

%%%%%%%%%%%%%%%%%%%%%%%%%%%%%%%%%%%%%%%%%%%%%%%%%%%%%%%%%%%%%%%%%%%%%%%%%%%
%%%%%%%%%%%%%%%%%%%%%%%%%%%%%%%%%%%%%%%%%%%%%%%%%%%%%%%%%%%%%%%%%%%%%%%%%%%
\section{Observations and point source photometry}
\label{obs}
%%%%%%%%%%%%%%%%%%%%%%%%%%%%%%%%%%%%%%%%%%%%%%%%%%%%%%%%%%%%%%%%%%%%%%%%%%%
%%%%%%%%%%%%%%%%%%%%%%%%%%%%%%%%%%%%%%%%%%%%%%%%%%%%%%%%%%%%%%%%%%%%%%%%%%%

The observations were acquired using the NIC2 camera of HST/NICMOS
under HST general observer programmes 7461 and 7880. For a description 
of the NICMOS instrument, see Thompson et al. (\cite{thompson}).
The observations for programme 7461 consisted of a total of 2560 s exposures 
in the F110W filter and 5120 s in F160W, and are described in \"Ostlin
(\cite{ostlin2000}) together with details on the reduction and data
analysis. 

The observations for programme 7880 were carried out during 10 orbits
and are summarised in Table {\ref{obstab}}. Due to the long wavelengths
of the filters, the observations were obtained by using ``chopping'',
i.e. separate sky frames were obtained in between each exposure on target. 
The observations employed the {\sc spiral-dith-chop} pattern in the {\sc multiaccum}
mode and the exposure sampling sequence used was {\sc mif512}. 
	Hence, for each orbit, five 512-second
exposures were obtained, three on the primary target and two on adjacent
sky positions. 
In order to improve the spatial sampling and photometric accuracy, small 
(1.5\arcsec) offsets were made in between the individual exposures for
both the target and sky positions.
Some images suffered from the so called pedestal effect, which was removed 
using the Pedestal Estimation and Quadrant Equalization Software, developed 
by R.P. van der Marel.

Before combining the target images, the background was subtracted using
the sky frames. Best results were obtained by first combining all sky
frames to a master sky which was then subtracted. The sky-subtracted target
frames were finally combined and rebinned to a finer pixel grid using 
the {\sc drizzle} package (Fruchter \& Hook, \cite{fruchter}).

Point source photometry was performed using the {\sc daophot} package (Stetson 
et al. \cite{stetson}). 
%The point spread function (PSF) was fitted over a radius corresponding
%to 3 undrizzled NIC2 pixels (or 0.225\arcsec). 
Aperture corrections to convert our {\sc daophot} magnitudes to total 
NICMOS magnitudes\footnote{NICMOS total magnitudes  are defined as the 
magnitude within an  0.5\arcsec  \  radius aperture plus $-0.156$ mag.; 
see www.stsci.edu/instruments/nicmos}
were determined by performing photometry of synthetic
NICMOS PSFs generated with the Tiny Tim software (Krist \& Hook 
\cite{krist}).

The observations were acquired at a different spacecraft roll angle 
compared to programme 7461, and the astrometry of point
sources for the two different programs were combined with the {\tt xy2rd}
task in {\sc stsdas}. In matching the photometry catalogues in different filters,
a tolerance of 0\arcsec.0375  was applied for the centre coordinates. 
Compared to \"Ostlin (2000), we detect somewhat fewer stars due to the 
narrower band passes of the F171M and F180M filters, and higher background
in F205W.
The zero points are given in Table \ref{obstab} and are in the {\sc Vegamag}
system. We will denote magnitudes in the different filters simply by 
their name, e.g. $F205W$.

\begin{figure*}
\includegraphics[clip=,width=0.5\textwidth]{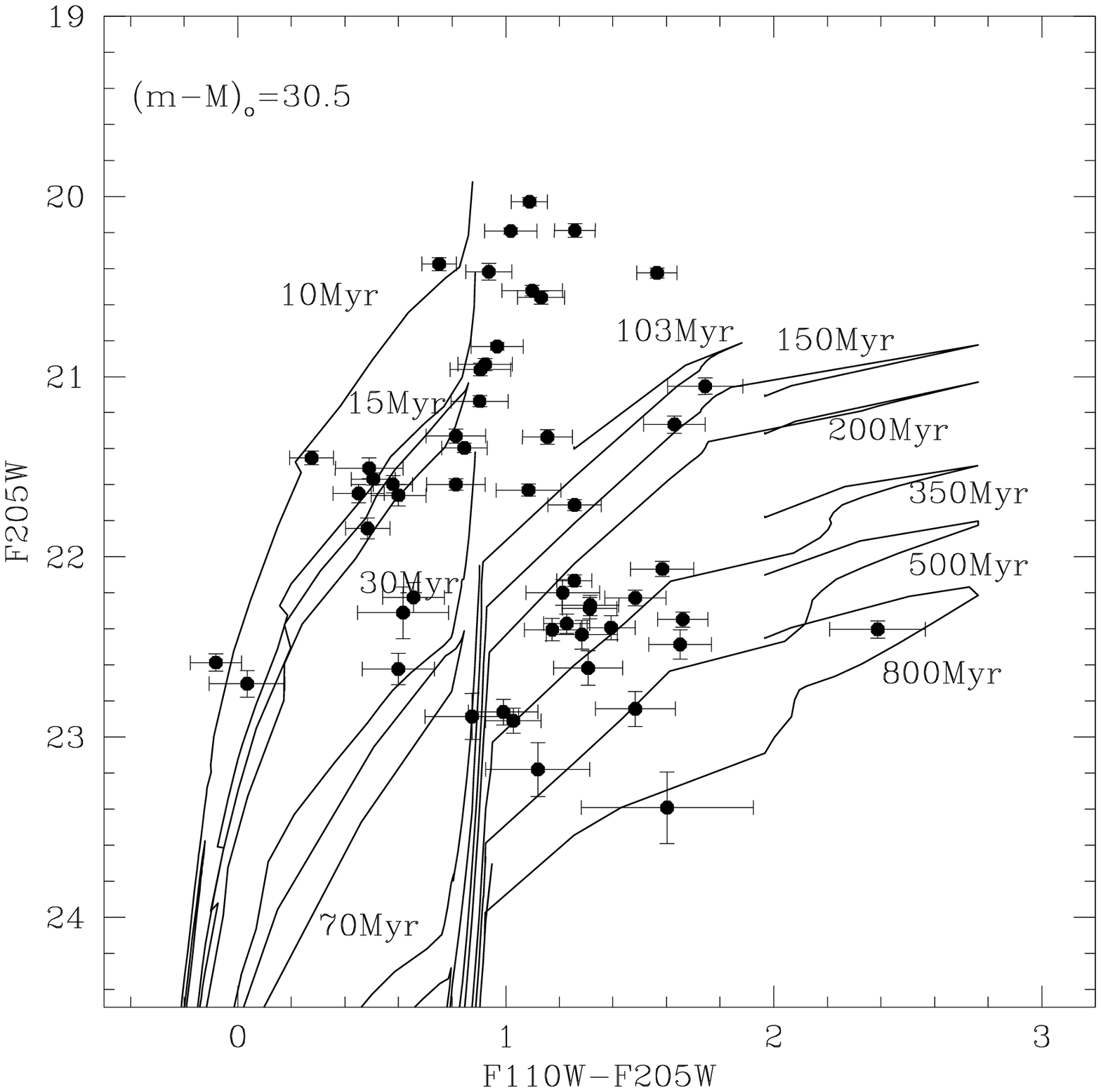}
\includegraphics[clip=,width=0.5\textwidth]{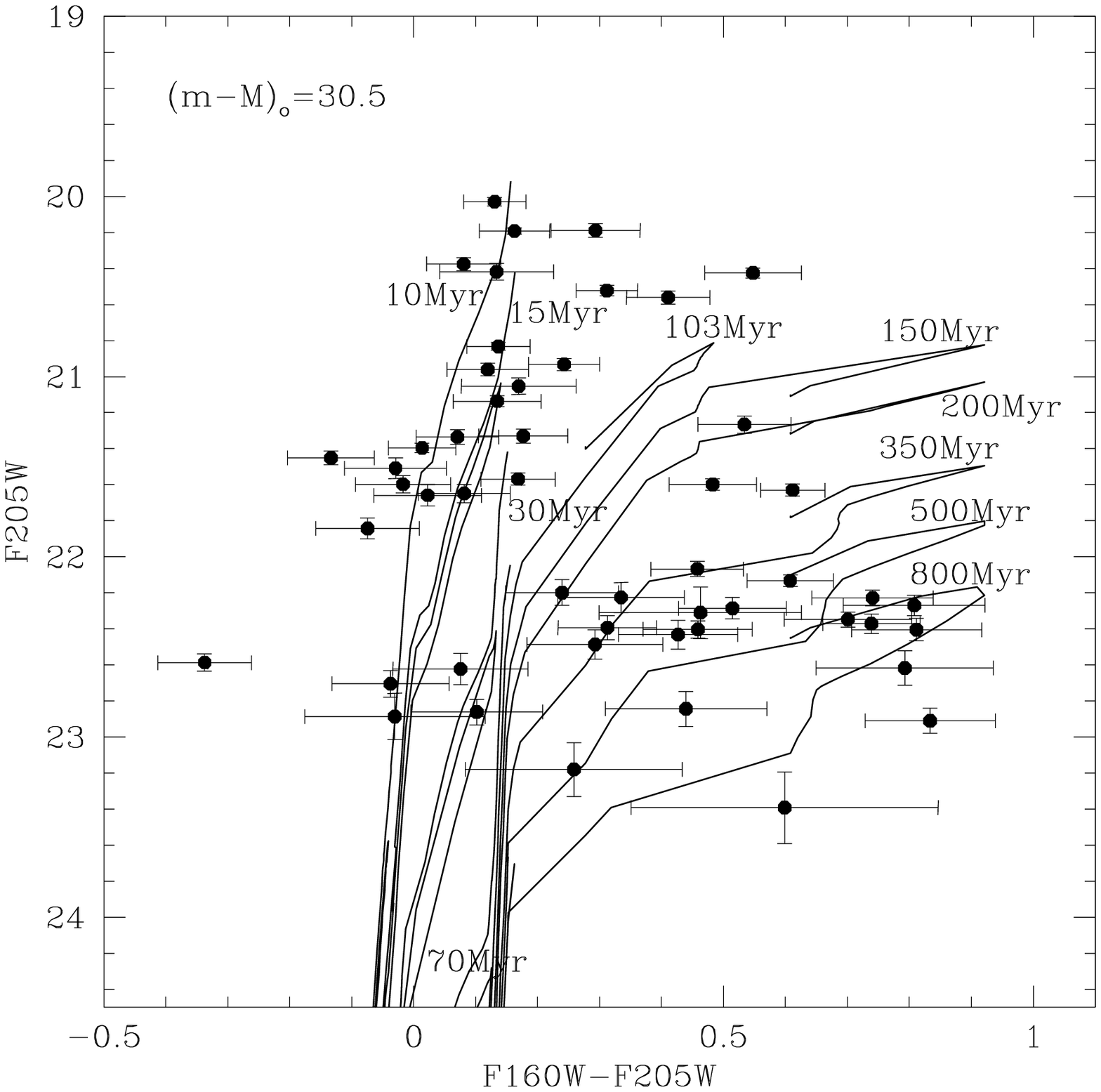}
\caption{Observed colour-magnitude diagrams: F205W vs. F110W--F205W 
(left) and F205W vs. F160W--F205W (right). Isochrones 
for a metallicity of $Z=0.0004$, and a distance modulus of 
$(m-M)_{\circ}=30.5$ are  overplotted. Each isochrone is labelled with its age.  }
\label{cmds}
\end{figure*}

%%%%%%%%%%%%%%%%%%%%%%%%%%%%%%%%%%%%%%%%%%%%%%%%%%%%%%%%%%%%%%%%%%%%%%%%%%
%%%
%%%	Surface photometry
%%%
%%%%%%%%%%%%%%%%%%%%%%%%%%%%%%%%%%%%%%%%%%%%%%%%%%%%%%%%%%%%%%%%%%%%%%%%%%

 \begin{table}[h]
\caption[]{Model and observed colours of extended emission features.
The model colours have been calculated with the Zackrisson et al. (2001) 
code and assume a burst with age 3 Myr, a standard  Salpeter IMF with 
a mass range of 0.08 to 120 \msun, and gaseous and stellar metallicities 
of $Z_{\rm stars}=0.001$ and $Z_{\rm gas}=0.0004$. Note that the effect 
of nebular emission on broad band colours sensitively depends on the assumed 
metallicity and the filter set used. The model colours presented here 
are only valid for I\,Zw\,18 and the NIC2 filter set.} 
\begin{flushleft}
\begin{tabular}{lllll} 
\hline
Colour &  \multicolumn{2}{l}{Model} & Filament & Blob  \cr
\cline{2-3}
 & stars  & stars+gas \cr
\noalign{\smallskip} 
\hline 
\noalign{\smallskip}
F110W--F160W & $-0.05$ & $0.23$  & $0.5$ & $0.3$ \cr
F110W--F205W & $-0.06$ & $1.13$  & $1.3$ & $0.9$ \cr
F160W--F205W & $-0.01$ & $0.90$  & $0.7$ & $0.6$ \cr
\hline
\end{tabular}
\end{flushleft}
\label{coltab}
\end{table}

\section{Extended and nebular emission}
\label{surf_phot}

The galaxy consists of two main star-forming complexes: The bright north-west 
(NW) and the more diffuse south-east (SE) regions, which are embedded in a faint 
irregular envelope. 
	When analysing the F205W ($\sim$K)  vs the F110W ($\sim$J) and F160W 
($\sim$H) images some features become evident (see Fig. \ref{image1}). 
To the west of the NW region there is a filament, and approximately
1\arcsec \  to the east of the centre of the NW complex there is a ``blob'' 
with size $\sim 0.5\arcsec$\ . Both features  are bright in the F205W 
filter but rather faint in both F110W and F160W. Their colours are given 
in  Table \ref{coltab}.

These features, and other larger filaments have also been 
seen in optical H$\alpha$ images (Dufour and Hester \cite{dufour}, Hunter and 
Thronson \cite{hunter}, \"Ostlin et al. \cite{ostlin1996}, Cannon et al. \cite{cannon}). 
The filament corresponds to the regions NW\,D4 and NW\,D5 in the nomenclature of 
Cannon et al. (\cite{cannon}), and the blob to NW\,1, all of which have very small 
reddening. 

The effect of nebular emission on the NICMOS colours  has been  computed for a 
stellar population including nebular emission using the models of 
Zackrisson et al. (2001). The realistic treatment of the nebular emission and 
the use of the HST+NIC2 throughput curves make these models suitable for our current 
purposes  (see \"Ostlin et al. 2003). 
We assumed a Salpeter initial mass function with mass range 0.08 to 120 \msun, 
$Z_{\rm stars}=0.001$, $Z_{\rm gas}=0.0004$, a hydrogen density of 100 cm$^{-3}$, a
filling factor of 0.1 and a gas covering factor of unity. Since the near-IR spectrum 
is completely dominated by  nebular emission  (line and  continuum)  for  ages 
less than a few Myr (see e.g. Bergvall \& \"Ostlin 2002) the composite
stellar+gaseous spectrum is a good approximation to a pure nebular spectrum. 
The model colours with and without gas emission for an age of $\le 3$ Myr and
constant star formation rate are
given in Table \ref{coltab}. 
Given the photometric uncertainties, the colours of the filament and the blob 
are consistent with pure nebular emission. 
The high H$\alpha$ surface brightness suggest that the blob is ionised from inside 
by a cluster of young stars, or alternatively could be a supernova remnant.
 	The effects of nebular emission on the colours for I\,Zw\,18 have also been 
discussed in Izotov et al. (2001), Papaderos et. al. (2002) and Hunt et al. (2003).

% Another possible contributor to the near IR spectrum is H$_2$ emission. Unless the 
% gas is thermally, rather than flourescently, excited one would expect it to be as
% strong in F110W as in the F205W filter (Black \& van Dishoeck 1987). 
% This suggests that we may be seeing a combination of H{\sc i} and
% H$_2$ emission from shocked gas, e.g. due to a past supernova explosion. 
% A last, but not very likely, possibility is that hot dust contributes to 
% the F205W flux.
%	The blob seen in Fig \ref{image1} is  also the source with the highest  
% $H\alpha$ surface brightness (Cannon et al. \cite{cannon}), and has virtually no
% internal reddening as derived from H$\alpha$/H$\beta$ photometry.
% The high H$\alpha$ surface brightness suggest that the blob is ionised 
% from inside by a cluster of young stars, and that 
% either H$_2$ emission from shocked gas, or a population of pre-main sequence stars 
% contribute to the colours. 
%	The blob has rather similar colours to the intermediate
% age red stars in I\,Zw\,18 (see Fig. \ref{colcol}) and could be consistent
% also with pre-main sequence stars stars since these are also red and
% luminous (Bernasconi and Maeder \cite{bernasconi}).

The possible contamination of point source photometry by patchy nebular emission 
has been assessed by comparing the spatial locations of our stars with the H$\alpha$
maps in Cannon et al. (\cite{cannon}). None of our carbon star candidates 
(see Sect. 4) with 
available F171M and F180M photometry are located near H$\alpha$ sources or 
regions with high  H$\alpha$
equivalent width. Of the AGB star candidates without F171M and F180M photometry 
that are classified solely based on the F110W, F160W and F205W colours, about five 
are located near, but not coincident with, H$\alpha$ sources in the SE complex.
Hence, we do not expect our point source photometry to be contaminated by 
nebular emission.
As discussed in Sect. 4.1, if contamination by nebular emission was an issue,
the effect would be that the intermediate-age population would be older than it
now appears.  Moreover, nebular gas with the metallicity of I\,Zw\,18 has
medium wide band colours $F171M-F160W \ge -0.2 $, and   $F180M-F171M \le 0$, 
incompatible with the colours of our carbon star candidates that we discuss below.

%%%%%%%%%%%%%%%%%%%%%%%%% pixel-pixel phot %%%%%%%%%%%%%%%%%%%%%%%%%%%%%%%%%%%%%%%%%

As described in Sect. \ref{pht}, carbon star candidates in I\,Zw\,18 are 
characterised by their location in the F171M--F160W vs F180M--F171M
colour--colour diagram. To test whether we could see any evidence for
carbon stars with luminosity below our limiting magnitude for PSF
photometry, we investigated the distribution of individual pixels in
these colours. This was done after binning the image to the original
pixel scale of NIC2 (0.075\arcsec) and applying a lower flux limit 
of $26.6$ magnitudes per binned pixel  in F171M and F180M, in order to 
omit too uncertain pixels. 
Pixels containing stars already described by the single star
PSF photometry were also omitted. In the crowded NW region, most pixels 
are contaminated with light from bright stars, but in the SE region
we see tentative evidence for carbon stars below our point source
detection limit (see Fig. \ref{surf_se} and Sect. \ref{se}).

%%%%%%%%%%%%%%%%%%%%%%%%%%%%%%%%%%%%%%%%%%%%%%%%%%%%%%%%%%%%%%%%%%%%%%%%%%%%
\section{Discussion: the nature of infrared-detected stars in I\,Zw\,18}
\label{pht}
%%%%%%%%%%%%%%%%%%%%%%%%%%%%%%%%%%%%%%%%%%%%%%%%%%%%%%%%%%%%%%%%%%%%%%%%%%%%

In the following subsections we will discuss the results of the resolved star
photometry. By comparing with model isochrones and empirical data of evolved
late type stars, we constrain the nature of of the faint red stars in I\,Zw\,18.

%%%%%%%%%%%%%%%%%%%%%%%%%%%%%%%%%%%%%%%%%%%%%%%%%%%%%%%%%%%%%%%%%%%%%%%%%%%
\subsection{Colour--magnitude diagrams and model isochrones}
\label{models}
%%%%%%%%%%%%%%%%%%%%%%%%%%%%%%%%%%%%%%%%%%%%%%%%%%%%%%%%%%%%%%%%%%%%%%%%%%%

In Figure \ref{cmds} we show the F205W vs. F110W--F205W, and F205W vs. F160W--F205W 
colour-magnitude diagrams (hereafter CMDs).  Similar to what was found by \"Ostlin 
(2000), the CMDs display two main features: The first feature consists of
bright (${\rm F205W\lsim\,22}$)  stars  with relatively blue colours 
(${\rm F110W-F205W\lsim\,1.2}$, ${\rm F160W-F205W\lsim\,0.3}$). 
The second feature consists of  faint (${\rm F205W\gsim\,22}$) stars
with red colours (${\rm F110W-F205W\gsim\,1}$ and 
${\rm F160W-F205W\gsim\,0.2}$). This 
suggests that the star formation history of I\,Zw\,18 has involved  two 
separate (in time) events or has been continuous over a longer time.

A means for interpreting the observed CMDs is provided by isochrones calculated 
from the stellar population models by Mouhcine (2002) and Mouhcine \& Lan\c{c}on 
(2002), which include the evolution of stars through the upper AGB phase. 
	The theoretical predictions
of these stellar population models have been extensively and successfully
compared to a variety of observational constraints. In order to constrain the  
stellar content of  I\,Zw\,18,  a set of isochrones with a
metallicity of $Z=0.02\,Z_{\odot}$ were calculated. For a detailed discussion 
of the models, their ingredients and limitations, the reader is referred 
to the papers cited above. An advantage of these models is the 
inclusion of carbon stars in a self-consistent manner, accounting for the 
dependence of mass loss and the sensitivity of 
carbon star formation to the stellar mass and abundance.
	This is of crucial 
relevance here as our aim is to identify carbon star candidates.  
To transform the theoretical luminosity--effective temperature diagram 
to a colour--magnitude diagram, we used the library of Lan\c{c}on \& 
Mouhcine (2002) for AGB stars, and  the compilation of Lejeune 
et al. (1998) for all other types of stars. 
The former library contains spectra of both oxygen-rich stars and carbon
stars. The 
isochrones have been calculated for the HST/NICMOS photometric system,  with 
the aid of the NIC2 throughput curves. 
%, allowing us to account for the impact of the spectral dichotomy 
%between oxygen- and carbon-rich stars on their spectrophotometric 
%properties.  

To compare the predicted isochrones to the observed colour-magnitude 
diagram, a distance to I\,Zw\,18 is needed. This is still a matter of debate,  
varying according to different authors from 10 Mpc to 20 Mpc. Hereafter, 
we will adopt a distance modulus of $(m-M)=30.5$ for I\,Zw\,18 (see \"Ostlin 
2000 for a discussion of the choice of this distance). Internal reddening 
by dust has been found to be very low (Cannon et al. \cite{cannon}), 
and has been neglected. As a word of caution we note that even if the average 
dust content is low, there are regions with a locally higher extinction 
($A_V < 0.5$ mag). 
	These are however not spatially coincident with the sources of primary 
interest in this paper and moreover, the effect on the infrared colour indices 
under study here are marginal (for instance $A_V=0.4$ mag corresponds to
$E(F160W-F205W)\sim0.02$).

Figure\,\ref{cmds} shows a set of isochrones, with ages noted in each panel, 
superposed on the observed CMDs. 
The isochrones show that once AGB stars appear in a stellar population,
at ages older than $\sim\,100\,$Myr, a red tail develops in the CMD 
and extends to quite extreme colours. As a consequence of high carbon star
formation efficiency at low metallicities, AGB stars turn into carbon 
stars almost immediately after they enter the red tail. The extent of 
the red tail, and particularly the carbon star sequence, depends on the 
calibration(s) used to transform the fundamental stellar properties of 
the AGB sub-population stellar to observed quantities. Using the stellar 
library of Lan\c{c}on \& Mouhcine (2002), the portion of isochrones with 
colour redder than ${\rm F110W-F205W}\simeq\,1.4$ and 
${\rm F160W-F205W}\simeq\,0.3$ are occupied by carbon stars.

For a distance modulus of $(m-M)=30.5$, the observed bright plume
with its relatively blue colours, is consistent with a young stellar population
in the age range  10-30\,Myr, corresponding to the age of turn-off stars 
with masses in the range of 21-10\,M$_{\odot}$. The fainter end of the 
this plume corresponds to a slightly older population, with age $\sim 50$
Myr and a turn-off   mass $\sim 6 M_{\odot}$.  
To account for the faint and red sequence, stars in the age interval 
100-800\,Myr are required. Changing the distance to the galaxy within  a 
reasonable range can not alleviate the need for stars older than 100\,Myr.
All the isochrones of stellar populations younger than a 100\,Myr predict
a plume of stars with colours bluer than ${\rm (F110W-F205W)\approx\,1}$ 
and ${\rm (F160W-F205W)\approx\,0.2}$. To occupy regions in the CMDs with
redder colours than these limits requires the presence of  intermediate-age 
stars with  turn off masses in the range  5--2\,$M_{\odot}$ evolving 
through the thermally pulsating AGB phase. This suggests that I\,Zw\,18  started 
forming stars at least 0.5\,Gyr prior to the current star formation event. 
Stars that are located in the CMD regions occupied by faint and red stars 
are, according to the isochrones, mostly carbon stars.

A comparison with the theoretical isochrones shows that while the 
intermediate-age stars in I\,Zw\,18  span a similar range in F160W--F205W
to the models
their F110W--F205W colours are on average bluer. 
This cannot be fully explained by the effect of a low initial metallicity 
of intermediate-age progenitors on the effective temperature, as this 
would make both colours bluer. 

One possibility is that the colours of 
intermediate-age stars have been contaminated by nebular emission in the neighbourhood. 
The effect of adding nebular emission is to make the  F110W--F160W colour appear bluer 
than its intrinsic value,
whereas the effect is much smaller for F160W--F205W since for this
colour index, nebular gas and AGB stars have similar colours. However, as
discussed in Sect. 3, only a small fraction of the stars are located near 
H$\alpha$ sources, and we do not believe that this could be the main mechanism,
but if it was, the net effect would be to make the intermediate-age population
in I\,Zw\,18 older than it now appears.

\begin{figure}
\includegraphics[clip=,width=0.5\textwidth]{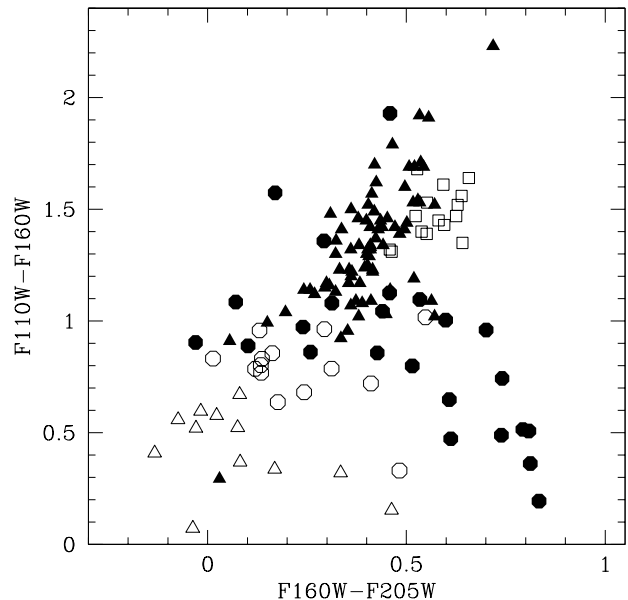}
\caption{F110W--F160W vs. F160W--F205W diagram for I\,Zw\,18 stars 
compared to a local sample of evolved stars. Also stars that were not 
detected in F171M and F180M are included. Solid circles show I\,Zw\,18 
intermediate-age stars, open circles show red (${\rm F110W}-{\rm F160W}>0.15$, 
and ${\rm F110W}-{\rm F205W}>1$) young stars and open triangles show bluer
young stars. The local sample is shown as solid triangles for oxygen-rich 
AGBs and supergiants, and open squares for carbon stars.}
\label{colcol}
\end{figure}

In figure\,\ref{colcol}  we compare the near-IR colours of stars in I\,Zw\,18,
to a local sample of evolved stars from Lan\c{c}on and Wood (\cite{lw}). 
%Intermediate-age stars in I\,Zw\,18 are shown as solid circles, while open
%circles and triangles show young stars. Open squares show local carbon
%stars, and solid triangles show local oxygen-rich stars.
A striking feature  of the intermediate-age stars in I\,Zw\,18 is that while they 
span a similar range in F160W--F205W, their F110W--F160W
colours are bluer compared to the local AGB star sample.
%, i.e. they rarely extend to similar red F110W--F160W colours as 
%observed for local carbon stars 
(see also Fig. \ref{cmds_cs}, and Sect. \ref{phot_prop} for more 
details.). The observed difference between the stars in I\,Zw\,18  and the local 
counterparts demonstrates that the low-metallicity stars  with intermediate 
mass and age, in I\,Zw\,18, have a F110W flux excess compared to
local stars. On the other hand, young stars in  I\,Zw\,18 show bluer colours 
than the local sample for both colours, and their location in the 
F110W--F160W vs. F160W--F205W diagram is a continuation of the local 
oxygen-rich star sequence, reflecting the known effect of metallicity on 
stellar effective temperature and broad-band photometry.

%%%%%%%%%%%%%%%%%%%%%%%%%%%%%%%%%%%%%%%%%%%%%%%%%%%%%%%%%%%%%%%%%%%%%%%%%%%%%%%%%%%%
\subsection{Photometric properties of metal-poor stars of intermediate age and mass }
\label{phot_prop}
%%%%%%%%%%%%%%%%%%%%%%%%%%%%%%%%%%%%%%%%%%%%%%%%%%%%%%%%%%%%%%%%%%%%%%%%%%%%%%%%%%%%

A caveat when modelling the near-IR CMDs of low-metallicity stellar 
systems, and for the identification of carbon star candidates, is 
the lack of representative  spectra for metal-poor evolved stars in 
the stellar library used to transform the  fundamental stellar parameters 
to observed quantities. The stellar library we have used is 
based on spectra of  nearby stars, and  is by construction biased toward 
metallicities consistent with the chemical evolution of our own Galaxy.
The observed difference in F110W--F205W colours between stars of intermediate age and mass
 in  I\,Zw\,18 and their local counterparts may imply that 
the carbon star sequence in metal-poor systems would not be tightly 
constrained by a direct comparison with isochrones based on stellar libraries 
of local stars. In a metal-poor environment 
 carbon stars are expected to appear at higher temperature compared 
to a metal-rich environment, and the progenitors of 
carbon stars will have higher effective temperature than their counterparts 
in the Large Magellanic Cloud (hereafter LMC) and the Milky Way. 
	In addition, the convective dredge-up
that enriches the stellar atmosphere with carbon, is triggered earlier and 
is more efficient at low metallicity (see Mouhcine \& Lan\c{c}on 2003 
for a detailed discussion of these issues).
However it is not an easy task to predict the distribution of 
carbon star colours. For Galactic and LMC carbon stars, 
$JHK$ colours become redder  as the  C$_2$ band strength, increase 
(Cohen et al. 1981). On the other 
hand, theoretical models show that the C$_2$ band head is sensitive to 
the carbon-to-oxygen ratio (Gautshy-Loidl 2001). Synthetic 
evolutionary models for AGB stars predict that the final carbon-to-oxygen ratio 
increases as the initial stellar metallicity decreases (Mouhcine 
\& Lan\c{c}on 2002, Marigo et al. 2002). Observationally, Matsuura et al. 
(2002) have shown that the equivalent widths of  molecular absorption 
bands  are systematically larger for carbon stars in the LMC than for carbon 
stars in the solar neighbourhood. They argue that the carbon-to-oxygen ratio 
is systematically larger in LMC carbon stars. 
The extent of line blanketing, responsible for the distinction between 
carbon stars and oxygen-rich stars, is related to the amount 
of carbon in the atmosphere. Thus, even if the progenitors of carbon stars 
 are hotter at low metallicities, their spectrophotometric properties 
might be unaffected by the lower metallicity as long as the opacity in the atmosphere 
is dominated by molecules based on carbon and other elements whose
 abundances are not drastically affected by the low initial 
metallicity. 

Red F110W--F160W colours of local carbon stars are due to the effect 
of molecular blanketing, i.e., CN and C$_2$ absorption bands. Atmospheric
carbon abundance is larger at low metallicity, which may enhance the
C$_2$ and CN absorption bands. On the other hand, as a consequence of 
the extremely low abundance of nitrogen in I\,Zw\,18 ($[{\rm N/O}]=-0.7, 
[{\rm N/C}]=-0.3$, in solar units, Izotov and Thuan 1999), the abundance 
of molecules that contain nitrogen may be lower in the atmosphere of 
AGB stars in I\,Zw\,18 than in their local counterparts, 
lowering the  CN-dominated opacity  in F110W filter, but increasing the 
opacity in both the F160W and F205W filter wavelength intervals, where  
C$_2$ dominates. This could give rise 
to the observed blue F110W--F160W and F110W--F205W
%, and the red  (F160W--F205W) 
colours. 
%
%The effect of a lower abundance on the other molecular species contributing 
%to the near-IR opacity and the effect on broad-band photometry is hard to 
%assess, but the red (F160W--F205W) colours could possibly also be related 
%to similar effects.
                                                                                 
The occurrence of  hot bottom burning in massive AGB stars, which convert 
a fraction of the newly dredged-up carbon into nitrogen via the CNO cycle, 
affects the abundance of both carbon and nitrogen in the atmosphere of such
stars. 
It would be hard to derive any meaningful constraints on the properties
of the hot bottom burning at low metallicity on the basis of the small
number of stars in our sample. 
Deeper observations of I\,Zw\,18, and/or other low-metallicity
systems, as well as detailed modelling of AGB stars, are needed to address this
question. 

These findings illustrate that I\,Zw\,18 constitutes an important laboratory
for understanding stellar physics and evolution at low metallicity. This
is not only true for AGB stars -- the discovery of Wolf-Rayet features in
I\,Zw\,18 a few years ago came as a surprise in view of the very low 
metallicity (Izotov et al. \cite{izotov1997}, Legrand et al.\cite{legrand},
Brown et al. \cite{brown}).

\begin{figure*}
\includegraphics[clip=,width=0.5\textwidth]{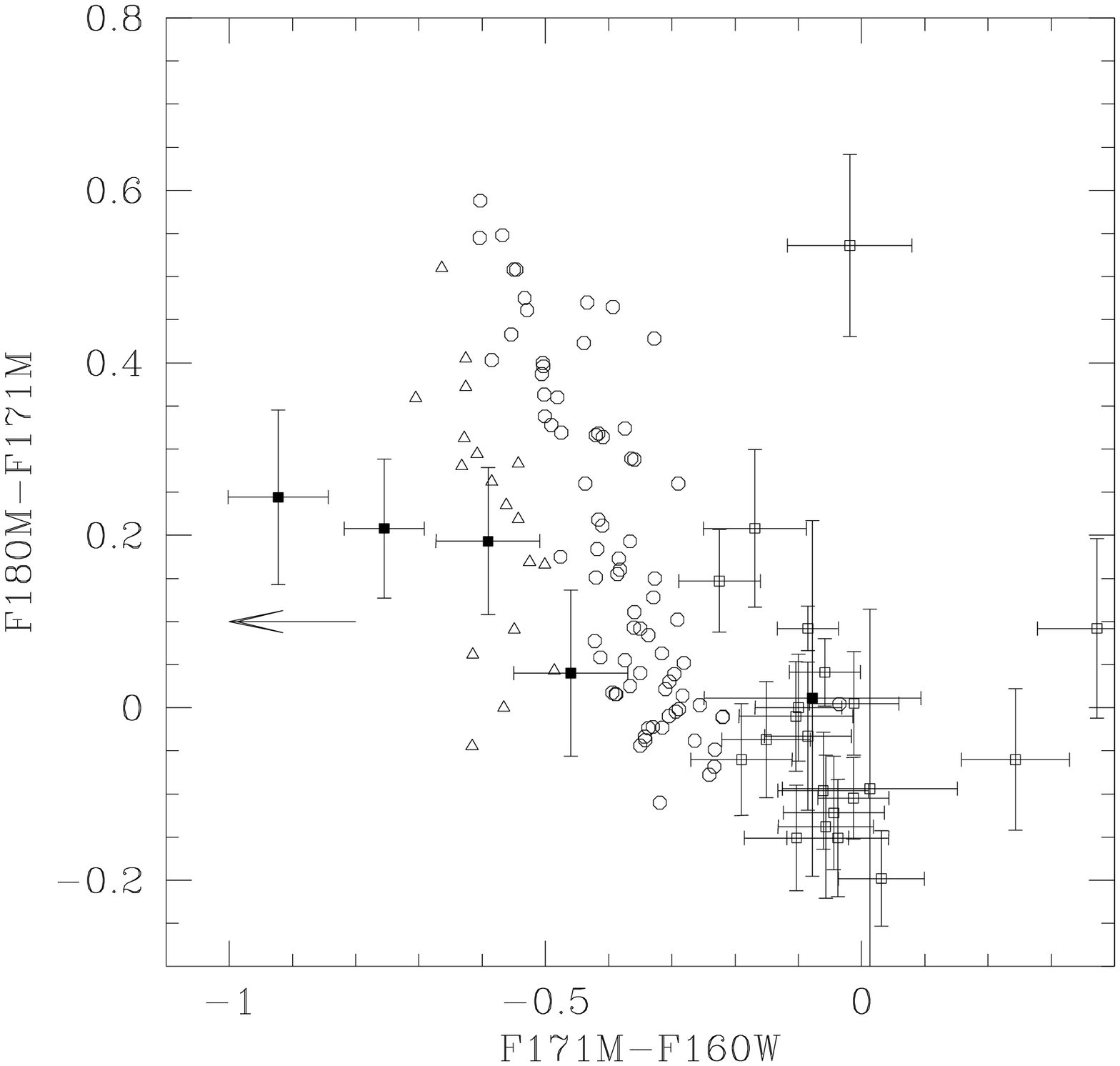}
\includegraphics[clip=,width=0.5\textwidth]{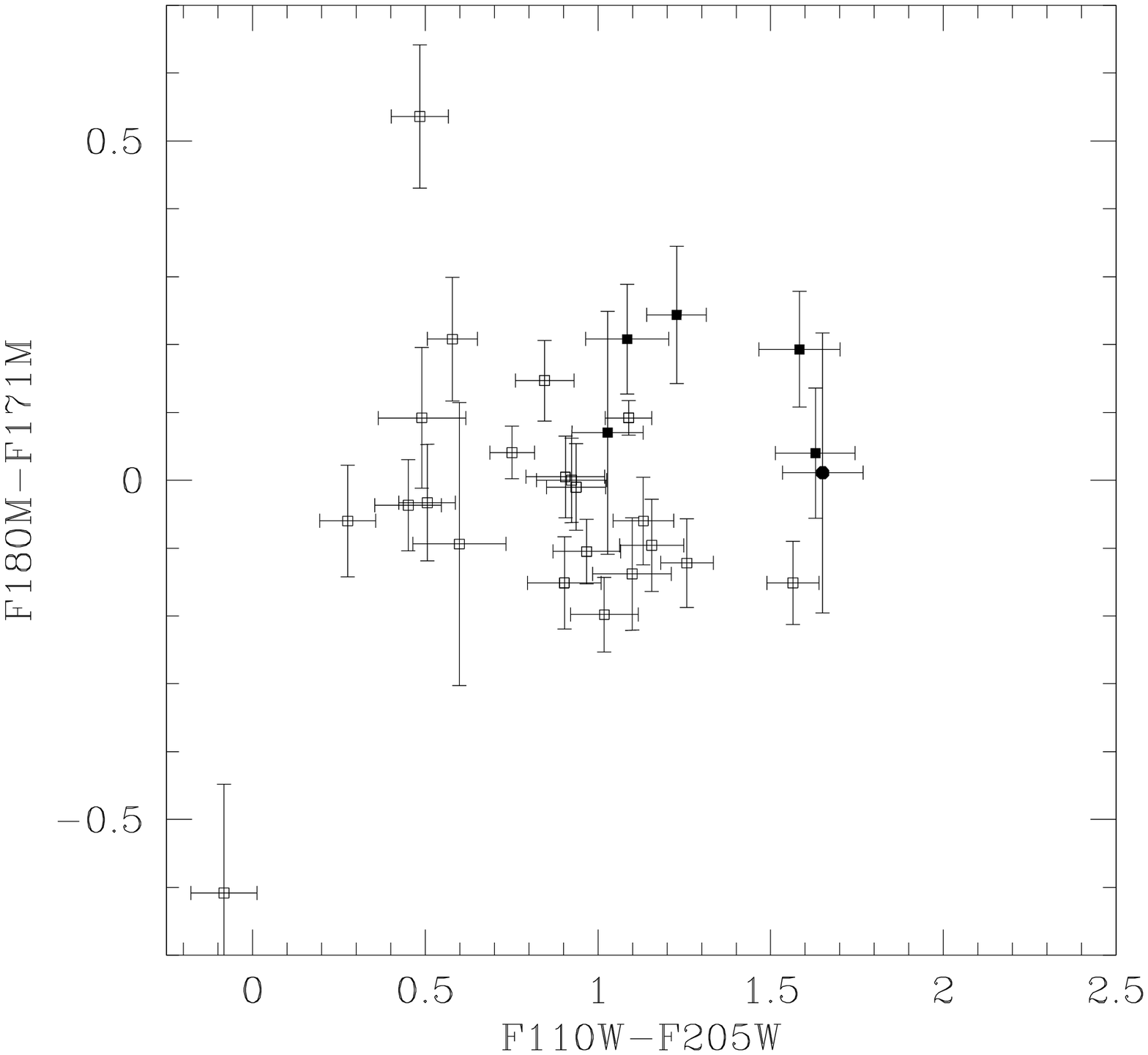}
\includegraphics[clip=,width=0.5\textwidth]{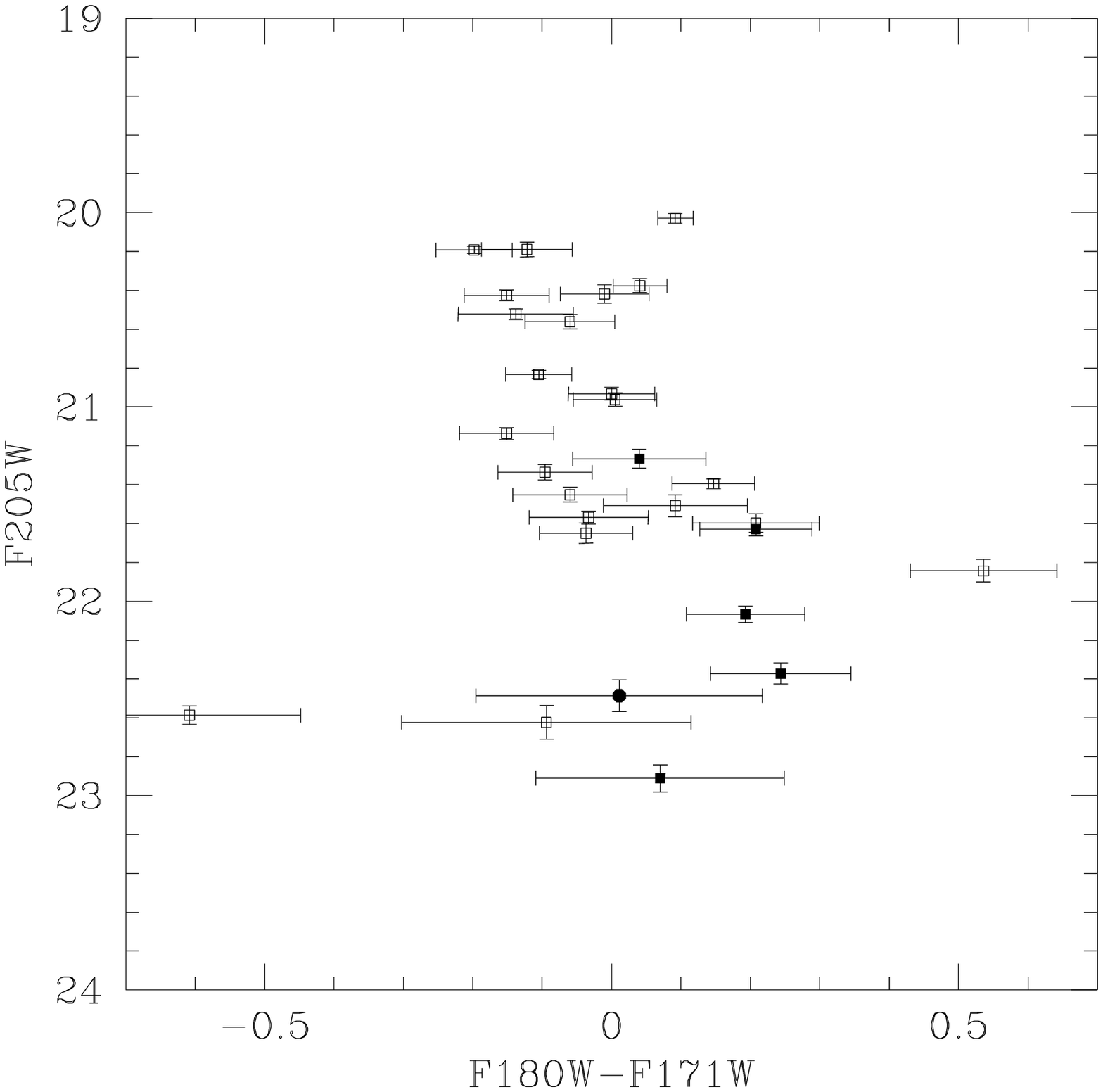}
\includegraphics[clip=,width=0.5\textwidth]{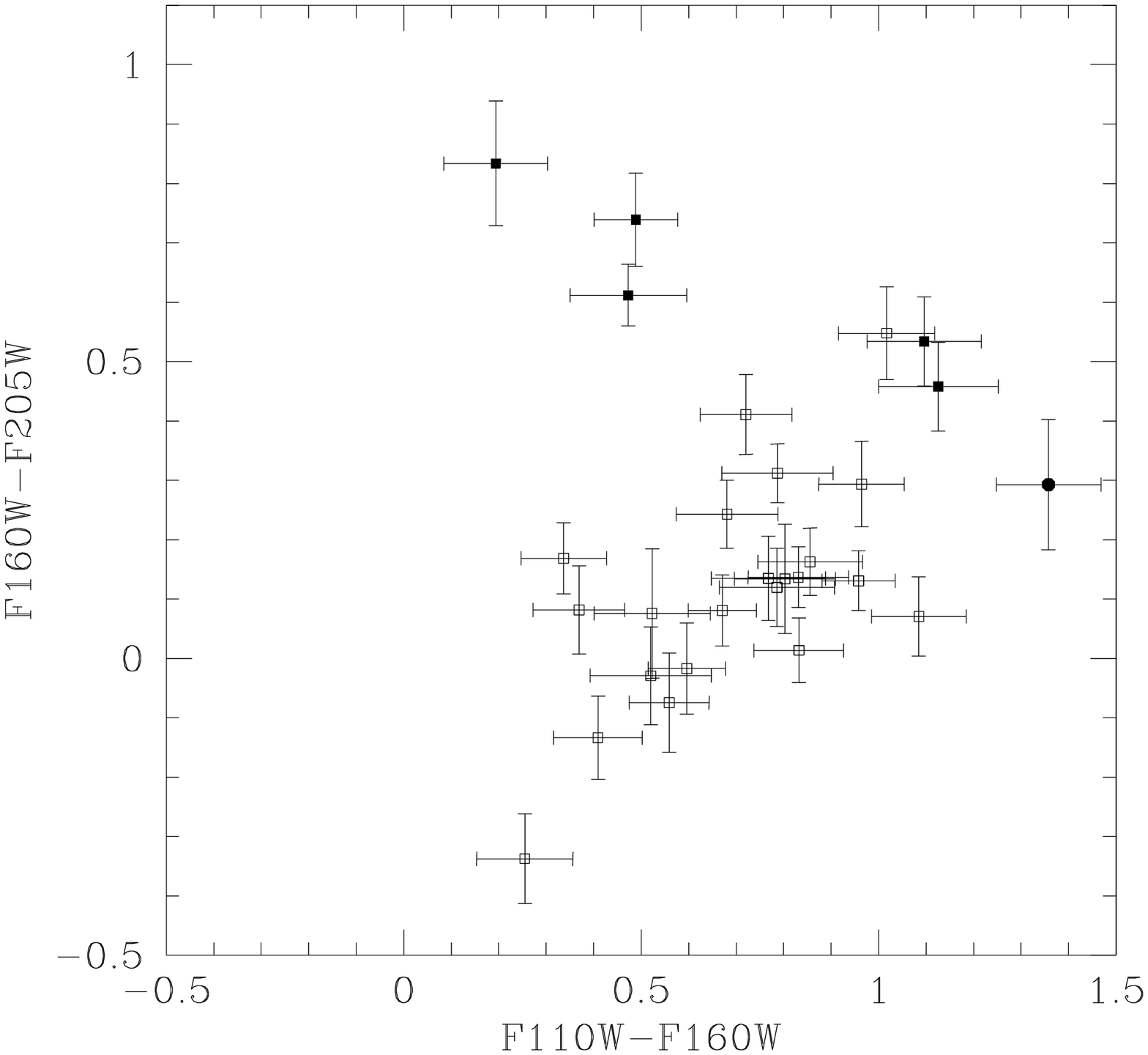}
\caption{Diagnostic diagrams for I\,Zw\,18 stars with available F171M
and F180M photometry. Upper left panel shows 
F180M--F171M vs. F171M--F160W compared to a local sample of evolved 
stars: triangles show spectroscopically classified carbon stars, and
circles show oxygen rich stars, giant stars, and supergiants. 
Filled squares show intermediate-age stars. The arrow indicates that one 
of these with ${\rm F171M-F160W}=-1.5$ is located to the left of the border 
of the plot. 
The upper right panel shows F180M--F171M vs. F110W--F205W diagram. 
The bottom left panel displays F205W vs. F180M--F171M diagram. The bottom
right panel shows F110W--F160W vs. F160W--F205W diagram. Stars 
classified as young stars are shown as open squares.
For the last three panels, carbon star candidates are shown as filled 
squares, an oxygen-rich intermediate-age star is shown as a filled 
circle.}
\label{diag}
\end{figure*}

\begin{figure*}
\includegraphics[clip=,width=0.5\textwidth]{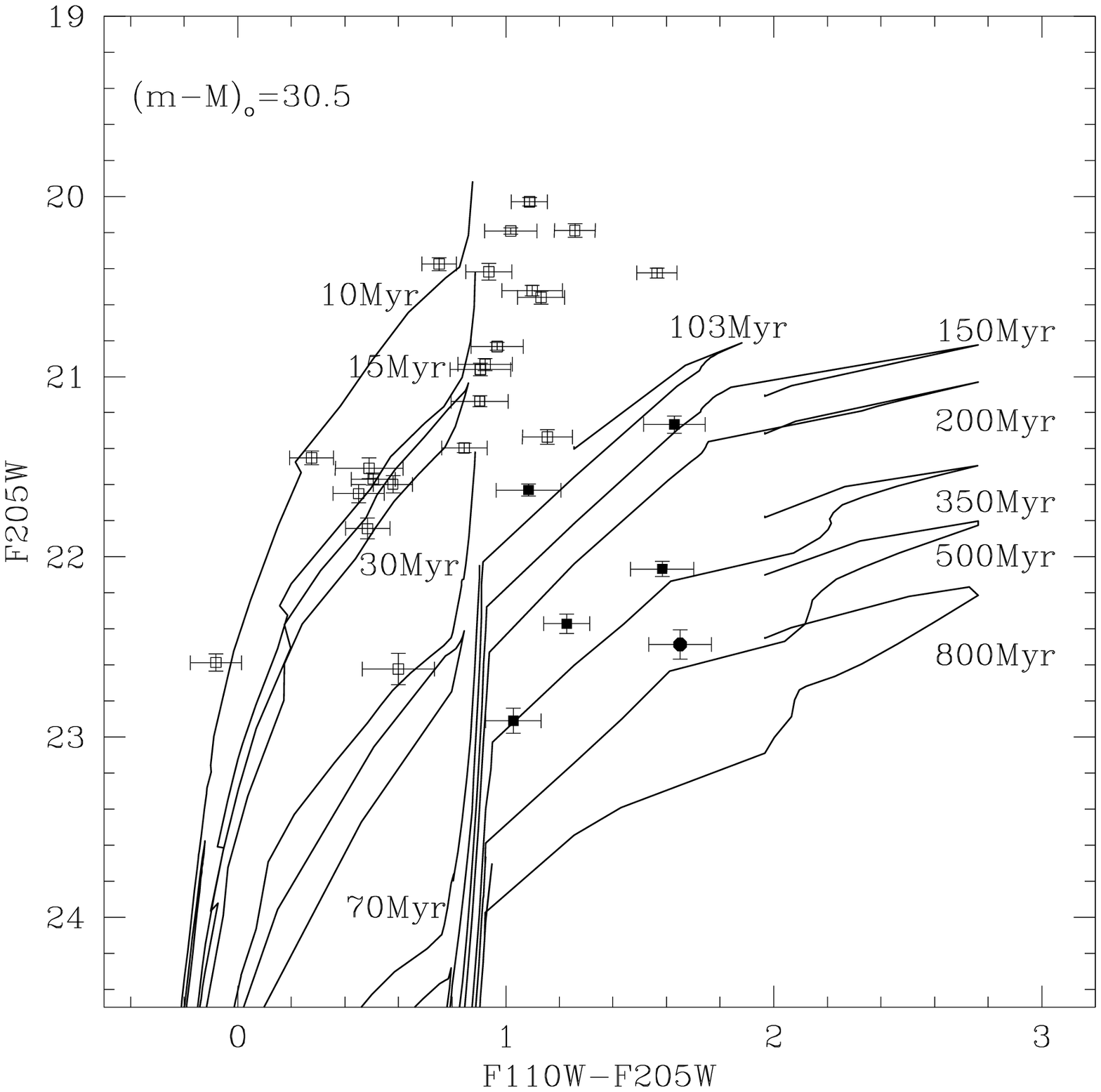}
\includegraphics[clip=,width=0.5\textwidth]{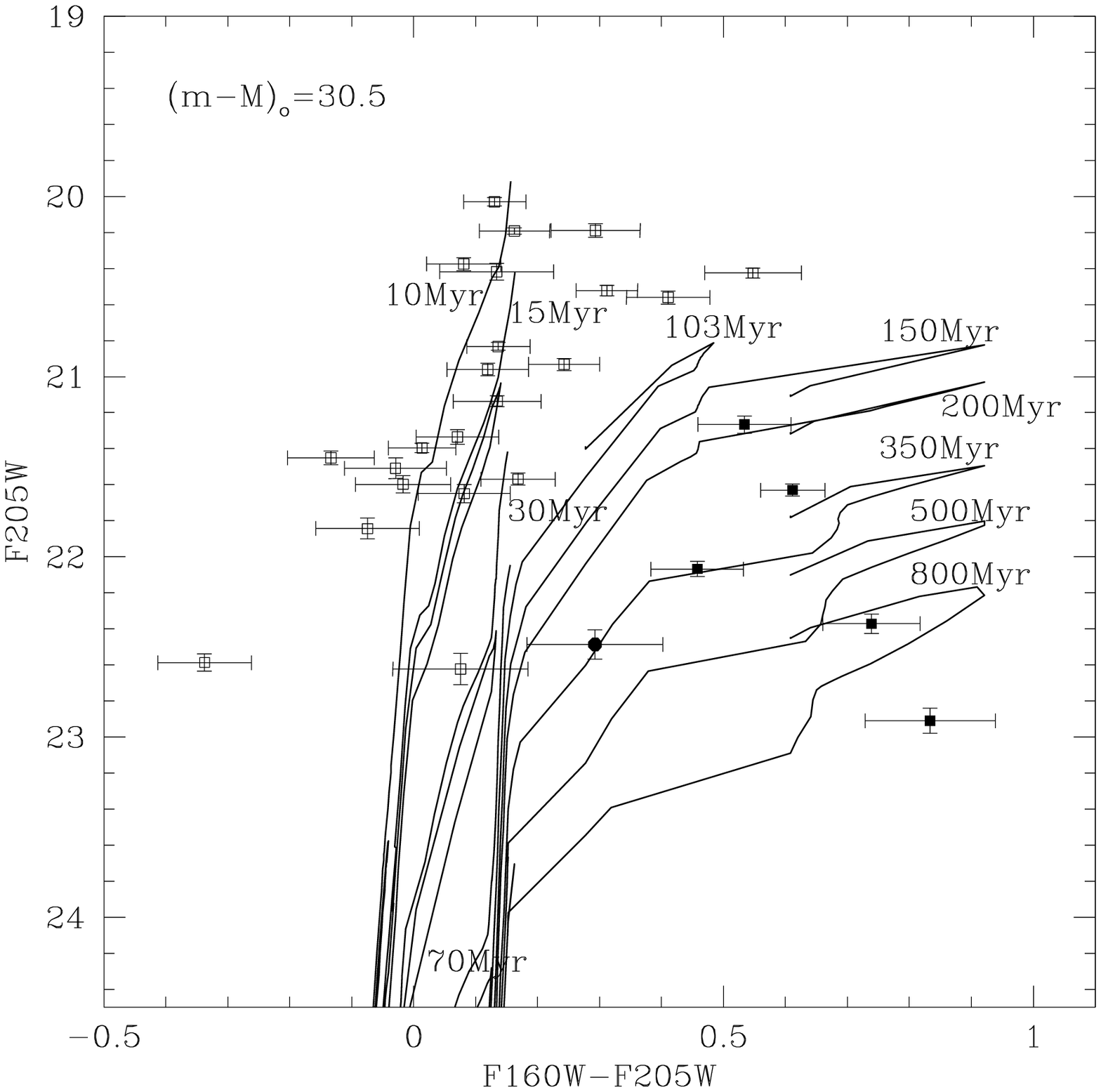}
\caption{Similar to Fig.\,\ref{cmds}, but restricted to stars with 
available medium-wide band photometry. Filled squares show carbon star 
candidates. The filled circle shows an oxygen-rich intermediate-age star 
(see text for more details).}
\label{cmds_cs}
\end{figure*}

\begin{figure*}
\resizebox{\hsize}{!}{\includegraphics{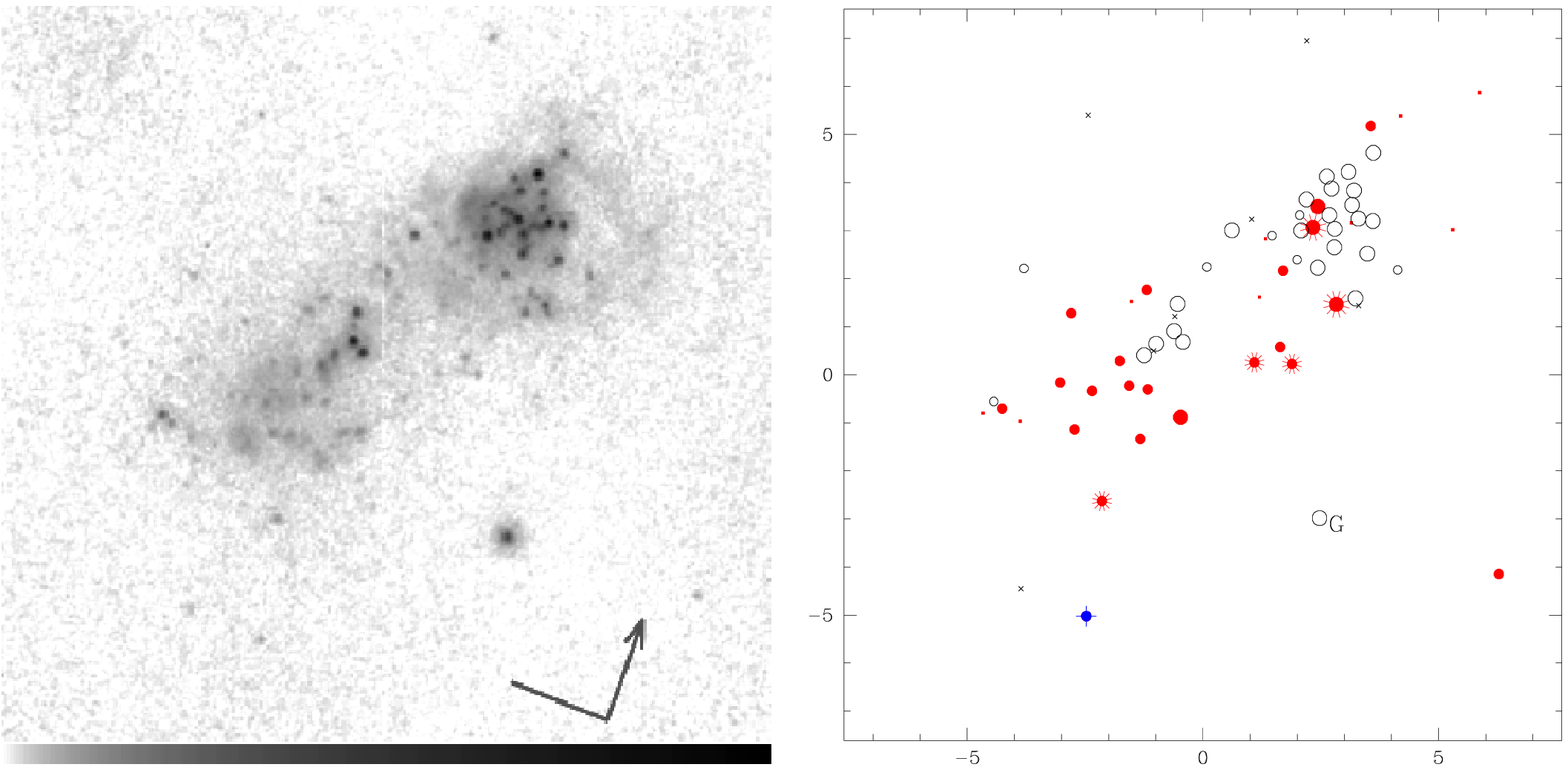}}
\caption{Spatial location of stars in I\,Zw\,18. 
{\bf Left:} co-added F110W, F160W and F205W frame with logarithmic 
intensity scaling. The orientation is indicated by the compass in the 
lower right corner, and the 
length of the arrow is $2\arcsec$. {\bf Right:} Filled (red) circles show intermediate-age 
stars detected in F110W, F160W and F205W, where those brighter than 
${\rm F205W}=22$ have slightly larger symbols. 
Those intermediate-age stars with available F171M and F180M photometry 
that have been identified as carbon star candidates have been highlighted 
by rays emanating from them. 
The candidate oxygen-rich star is marked with a plus (and shown in blue). 
Open circles show young ($<100$ Myr) stars, where stars brighter than 
F205W$=22$ are shown by larger symbols. 
Stars detected in F110W and F160W only are shown as small crosses, if 
F110W--F160W$<0.7$, or as small dots otherwise. One source suspected to be 
a background galaxy is in addition labelled with a ``G''.}
\label{st_dist}
\end{figure*}

%%%%%%%%%%%%%%%%%%%%%%%%%%%%%%%%%%%%%%%%%%%%%%%%%%%%%%%%%%%%%%%%%%%%%%
\subsection{Carbon star classification through medium-wide band photometry}
\label{narrow}
%%%%%%%%%%%%%%%%%%%%%%%%%%%%%%%%%%%%%%%%%%%%%%%%%%%%%%%%%%%%%%%%%%%%%%

An alternative empirical approach to determine the nature of intermediate-age 
stars is to compare  equivalent data for stars with known classification. 
In this subsection we try to classify  the stars in I\,Zw\,18 by comparing
their photometric properties to those of a local sample from Lan\c{c}on \& 
Wood (2000) that includes AGB stars (both oxygen-rich stars and carbon
stars), red giants, and supergiants. 
% Carbon star candidates are defined as stars that occupy similar locations 
% in the colour-colour diagrams as local carbon stars. 

The F180M--F171M index is a measure of the C$_2$ band strength for carbon 
stars {\it or} the H$_2$O  band strength for oxygen rich stars. It is able to separate 
between evolved AGB stars and other types of evolved stars, but 
it cannot separate  the two sub-populations of AGB stars 
(Alvarez et al. 2000). 
% The NICMOS medium-wide filters are wide enough 
% to be sensitive to the broad absorption bands.
 Fortunately, combining this index with other information can do a good 
job in separating both AGB sub-populations. 
Indeed, local  carbon stars split from the rest of the local 
sample in the F180M--F171M vs. F171M--F160W diagram. The F171M--F160W
index measures the slope of the spectrum between the centre and the edge
of the F160W filter. It has a numerically larger 
 for oxygen-rich stars than for 
carbon stars as a consequence of the presence of broad H$_2$O absorption 
band, while for carbon stars the flux at the edge  of the filter is larger 
than in the centre. As the spectral features of carbon stars are not
strongly sensitive to stellar temperature (Lan\c{c}on \& Mouhcine 2002), 
the behaviour of carbon stars in the (F180M--F171M) vs. (F171M--F160W) 
diagram should in principle  be independent of metallicity.
To identify carbon star candidates, we consider stars that are classified 
as intermediate-age stars in the near-IR CMDs (Fig. 2), i.e., older than 
100\,Myr. Six of these have available F180M and F171M photometry.

In the upper left panel of Fig.\,\ref{diag}, filled squares show the 
distribution of intermediate-age I\,Zw\,18 stars in the 
F180M--F171M vs. F171M--F160W diagram,  compared to the Lan\c{c}on 
\& Wood (2000) local sample, where triangles show carbon stars, and 
circles show all other oxygen rich stellar types, i.e. giants, supergiants, 
and oxygen-rich AGB stars. 
As the diagram shows, the combination of F180M--F171M and F171M--F160W 
indices can efficiently separate spectroscopically identified oxygen-rich 
and carbon stars, as the carbon stars have systematically smaller F171M--F160W
at a given F180M--F171M.
The six intermediate-age stars in I\,Zw\,18  fall into two groups in this 
diagram: Five stars, generally with positive F180M--F171M index, fall 
close to, or to the left of, the local carbon stars.  Only one intermediate-age 
star in I\,Zw\,18 falls close to the location occupied by local supergiant 
and oxygen-rich AGB stars. The other I\,Zw\,18 stars  in this region have 
too blue broad band colours and/or are too bright to be classified as 
intermediate-age stars, although there are two borderline cases.
 Bright stars 
with ${\rm F110W-F205W\lsim\,1}$ tend to have ${\rm F171W-F160W\sim\,0}$,
while fainter stars span a larger range in F171W--F160W. There 
seems to be a anti-correlation between the F180M--F171M and F171M--F160W 
indices for intermediate-age I\,Zw\,18 stars: the former index gets stronger 
on average as the latter index get weaker, and the same trend is seen for
local/LMC carbon stars. Hence, according to the upper left panel of 
Fig.\,\ref{diag}, the  stars 
in I\,Zw\,18 that are located in the region occupied by local carbon stars 
are likely carbon stars. The single  intermediate-age star located close to 
oxygen-rich stars is likely an  oxygen-rich AGB star. Below, we use additional 
diagnostic diagrams to confirm this classification. Note that, as discussed in 
Sect. 3, this classification is completely insensitive to any possible 
contamination of nebular emission.

The upper right panel of Figure\,\ref{diag} shows the 
F180M--F171M vs. F110W--F205W diagram for stars in I\,Zw\,18.
%, with carbon star candidates shown as filled squares and the
% oxygen-rich AGB star candidate as a filled circle. Young stars are shown
% as open squares. 
The figure shows again the dichotomy between  intermediate-age/mass stars and 
young/massive stars. For stars redder than ${\rm F110W-F205W\approx\,1}$, 
intermediate-mass stars tend to have positive F180M--F171M index, while 
supergiant stars have in average negative F180M--F171M index. 
Only two supergiants (with $M_{\rm F205W}<-9$) have a positive 
F180M--F171M index, including the most luminous star in the sample, and 
of all the young stars, only 6 out of 22 have ${\rm F180M-F171M>0}$.
The figure shows also the inability of the F180M--F171M index alone, 
to separate between oxygen-rich and carbon stars. The 180M--171M 
index, as expected for intermediate-mass stars, gets stronger on average as 
the stars get fainter as shown in the bottom left panel of Fig.\,\ref{diag}.

In the bottom right panel of Fig.\,\ref{diag}, we show that carbon 
star candidates tend to have redder F160W--F205W colour at a given 
F110W--F160W colour, similar to what is seen for LMC/Galactic carbon 
stars (Fig. \ref{colcol}). In principle, nebular contamination would give
a similar behaviour, but this explanation is unlikely since the stars
with the most deviant colours are the carbon star candidates which are not
located near any H$\alpha$ sources. 
	This rather supports the molecular blanketing interpretation, 
as opposed to a simple effective temperature effect. Other stars that 
show slightly redder F160W--F205W colours at a given F110W--F160W, but 
not classified as carbon stars, are too blue and/or too luminous to be  
intermediate-mass stars. Fig.\,\ref{cmds_cs} shows the same CMDs as 
in Fig.\,\ref{cmds}, but restricted to stars that have been detected in 
medium-wide filters. It is clear that almost all 
stars older than $\sim\,100\,$Myr are carbon stars. This is consistent with 
the theoretical models that predict that carbon stars dominate 
intermediate-age stellar populations at low metallicities.

We have shown that the F171M--F160W vs. F180M--F171M diagram can
separate probable carbon stars from other late type stars. 
If we relax the requirement of having photometry in both the F171M 
and F180M filters, and look also at stars accurately detected in F171M only,
the sample is increased with four more stars. 
Two of these are faint and blue and most probably young stars, whereas
two have broad band photometry consistent with intermediate-age stars. 
The two latter stars have F171M--F160W $= -0.1$ and $-0.3$, hence
most consistent with local O-rich stars. However, without F180M--F171M
photometry, this classification becomes tentative. Whereas our five carbon 
star candidates separate well from the rest of the stars in a F171M--F160W
vs. F160W--F205W diagram this is not true for carbon stars in general as 
can be understood from the upper left of Fig. \ref{diag}.

%%%%%%%%%%%%%%%%%%%%%%%%%%%%%%%%%%%%%%%%%%%%%%%%%%%%%%%%%%%%%%%%%%%%%%%%%%%%%
\subsection{Spatial location of stars in I\,Zw\,18}
\label{spat}
%%%%%%%%%%%%%%%%%%%%%%%%%%%%%%%%%%%%%%%%%%%%%%%%%%%%%%%%%%%%%%%%%%%%%%%%%%%%%

The spatial location of both young supergiant stars and intermediate-age 
stars are shown in Fig. \ref{st_dist}. Open circles show young stars, 
and filled circles show intermediate-age stars. Carbon stars candidates 
and the candidate AGB oxygen-rich star are highlighted. 
The figure shows that luminous supergiants are concentrated to the 
NW  and  SE regions, where the star formation 
is active, while AGB stars are more uniformly distributed.
 
\begin{figure*}
\resizebox{\hsize}{!}{\includegraphics{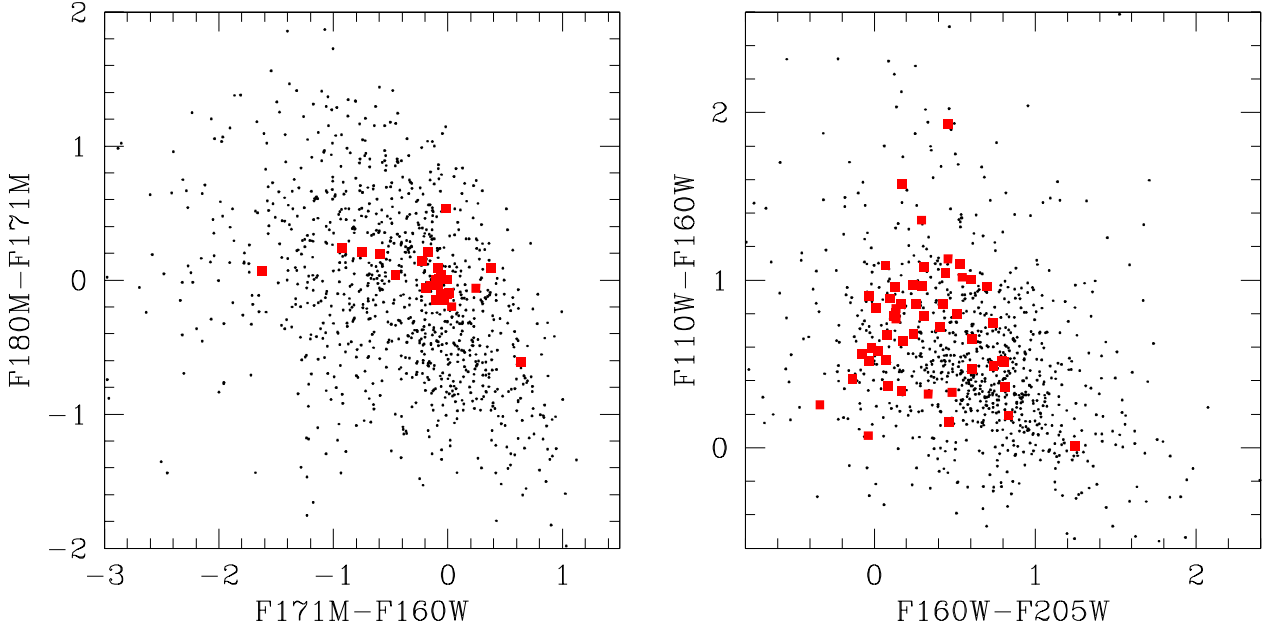}}
\caption{Pixel photometry of south-east region 
{\bf Left:} F171M--F160W vs F180M--F171M photometry. Each dot shows the 
photometry of individual pixels (binned to the original 0.075 arcsec/pixel 
scale of NIC2) in the south-east region with a luminosity in F171M and F180M 
brighter than 26.6 magnitudes, and excluding the point sources detected these 
filters. The filled squares show the corresponding photometry for stars in 
I\,Zw\,18 (see the upper left panel of Fig. \ref{diag}).
{\bf Right:} F160W--F205W vs F110W--F160W photometry, for the same pixels 
as presented in the left plot. The filled squares show all stars in I\,Zw\,18 
(see Fig. \ref{colcol})}
\label{surf_se}
\end{figure*}

%%%%%%%%%%%%%%%%%%%%%%%%%% SURF PHOT SE-REGION %%%%%%%%%%%%%%%%%%%%%%%%%%%%%

\subsection{Pixel photometry of unresolved stars}
\label{se}

As described in Sect. \ref{surf_phot} we have analysed the colours 
of individual pixels in the SE region. The result for a 30 arcsec$^2$ 
area is shown in Fig. \ref{surf_se} (noisy pixels and pixels contaminated
by detected stars have been excluded, for details see Sect. \ref{surf_phot}), 
together with the photometry of the stars discussed above. 
The region is approximately quadratic with a centre at  $(x,y) = (-2,0)$ in
the coordinate system of Fig. \ref{st_dist}.
%
%Clearly, for both the (F171M--F160W) vs (F180M--F171M)
%and the (F160W--F110W) vs (F205W--F160W) colour-colour diagrams, the pixel
%photometry reinforce the trends seen for the individual star photometry.	
%In addition to the general trend followed by stars in I\,Zw\,18 and 
%our galaxy (see Fig. \ref{diag}) there is an excess of pixels that 
%occupy the location of spectroscopically identified carbon stars. 
%

We have seen that the F171M--F160W vs F180M--F171M colour-colour diagram is an 
efficient tool for discriminating carbon stars and other class of objects.
Pixels that occupy the location of spectroscopically identified carbon 
stars in this diagram are then considered to be dominated by carbon stars.
	Figure 7 shows that individual pixels have a broad range of colours
indicating that the unresolved stellar population contains both young
and intermediate-age stars. Interestingly, the carbon star region
(see upper left of Fig. 4) is well populated.
The integrated colours for pixels in the carbon star region are F171M--F160W
$=-1.1$ and F180M--F171M$=0.0$, and have a total F160W magnitude of $21.6$. 
For single stars 
detected in all 5 passbands, the limiting magnitude is F160W $\approx 23.5$ 
meaning that the unresolved pixels could correspond to, on the order of 
10 carbon stars with F160W$\sim 24$, or $M_{\rm F160W} = -6.5$.

Taking the area of I\,Zw\,18 to be $14\arcsec \times 8\arcsec$ (the angular 
size at $\mu_R = 25$ mag/arcsec$^2$, Papaderos et al. 
\cite{papaderos}), the surface density of detected carbon stars in I\,Zw\,18 
as a whole is $\sim 15$ per kpc$^2$. The SE region has an area which is $27\%$ 
of the total assumed area and contains one of the 5 identified carbon stars.
If the SE region is assumed to be representative for the total AGB population 
of I\,Zw\,18, the pixel photometry of this region suggests that the surface 
density may be on the order of 100 per kpc$^2$, and the total number of 
carbon stars on the order of 50, but of course this sensitively depends on 
the assumed luminosity function of carbon stars. This would mean that 
I\,Zw\,18 has a carbon star surface density rather typical for faint 
metal-poor dwarf galaxies (see Groenewegen \cite{groenewegen}).
The absolute magnitude of I\,Zw\,18 is $M_V \approx -14$, but this is 
boosted by the presence of many young stars.
Assuming that the average metallicity of I\,Zw\,18 stellar content is 
similar to its gas phase metallicity, i.e., $Z=0.02\,Z_{\odot}$, I\,Zw\,18 
falls on the correlations of the total number of carbon stars against galaxy 
luminosity, and the normalised number of carbon stars to galaxy luminosity 
against galaxy metallicity respectively, as defined by Local Group galaxies (Groenewegen \cite{groenewegen}). Consequently carbon star content of 
I\,Zw\,18 appears to be similar to these observed for metal-poor Local Group 
dwarf galaxies. 
Given the significant difference between the luminosity-weighted ages of 
the Local Group dwarf galaxies and I\,Zw\,18, this may indicate that the 
scaling correlations that regulate the properties of carbon star content 
in galaxies could be metallicity rather than age.

\section{Conclusions}
\label{concl}

We have presented new broad and medium-wide band HST/NICMOS imaging of the 
very metal-poor blue compact galaxy I\,Zw\,18. The data indicate that an 
intermediate-age stellar component is indeed present, and that the galaxy
was not formed during the last 100 Myr. 
%The ongoing star formation occurs within an extended body of older stars. 
We emphasise the usefulness of combined broad- and medium-wide band 
near-IR photometry for detecting and identifying evolved intermediate-age 
stars whose energy distribution peak in the near-IR.

Based on comparison with isochrones and on empirical matching of stars in 
I\,Zw\,18 to morphological features of known stellar populations in different 
diagnostic diagrams we have identified carbon star candidates. 
	Out of the  stars classified as intermediate-age 
through F110W ($\sim$J), F160W ($\sim$H), 
and F205W ($\sim$K) photometry, six have available  medium-wide band 
photometry in the F171M and F180M filters, and five of these are probable 
carbon stars. This suggests that the bright AGB star population in I\,Zw\,18 
is dominated by carbon stars. At a distance of more than 10 Mpc, these are 
the most distant resolved carbon stars discovered as yet. 
 	Investigating the broad and medium-wide band photometry for pixels 
in the south-east region we find support for more carbon stars lurking behind 
the single star detection threshold. 

We find that intermediate-age stars in I\,Zw\,18 show relatively blue 
F110W--F160W colours compared to local AGB stars. The opacity source in 
AGB stars for the wavelength region covered by the F110W filter is dominated 
by the CN molecule, and we speculate that the blue colours may be due to the 
very low nitrogen abundance in I\,Zw\,18.

\begin{acknowledgements}
We are indebted to A. Lan\c{c}on for very useful and enlightening discussions, 
and to E. Zackrisson and N. Bergvall for computing a model with nebular
emission for IZw18 and adopting it to the HST/NICMOS filter system.  
H. Olofsson is thanked for comments on a draft manuscript, and G. Olofsson and 
K. Eriksson for useful discussions. Thanks also to S. Nasoudi-Shoar for some 
practical assistance, and to R. Cumming for help with  improving the language.
 Finally, the referee is 
thanked for a thorough report with many useful suggestions. 
In this paper, we have made use of the Pedestal Estimation and Quadrant 
Equalization Software developed by Roeland  P. van der Marel. 
This work was supported by the Swedish Research Council and the Swedish 
National Space Board.

\end{acknowledgements}

\end{document}